# Liu Estimator in the Multinomial Logistic Regression Model

Yasin Asar[1*] and Murat Erişoğlu[2]

[1] Department of Mathematics and Computer Sciences, Necmettin Erbakan University,
Konya, 42090, Turkey
yasar@erbakan.edu.tr, yasinasar@hotmail.com
*Corresponding Author

[2] Department of Statistics, Necmettin Erbakan University, Konya, 42090, Turkey
merisoglu@erbakan.edu.tr

## Abstract

This paper considers the Liu estimator in the multinomial logistic regression model. We propose some different estimators of the biasing parameter. The mean square error (MSE) is considered as the performance criterion. In order to compare the performance of the estimators, we performed a Monte Carlo simulation study. According to the results of the simulation study, we found that increasing the correlation between the independent variables and the number of regressors has a negative effect on the MSE. However, when the sample size increases the MSE decreases even when the correlation between the independent variables is large. Based on the minimum MSE criterion some useful estimators for estimating the biasing parameter d are recommended for the practitioners.

# 1. Introduction

In linear regression, the multicollinearity problem occurs when the explanatory variables are intercorrelated. When this problem occurs in multiple linear regression, it is not convenient to estimate the parameters by using the ordinary least squared (OLS) method due to the large variance and the instability of the OLS method. Similarly, in binary and multinomial logistic regression models the maximum likelihood estimation (MLE) method has these drawbacks. Thus, the inferences based on the regression analysis may not be trustworthy.

Hoerl and Kennard (1970) suggested the ridge regression to solve this problem in the multiple linear regression model. In literature, many researchers have proposed different ridge estimators. To mention a few, Kibria (2003), Khalaf and Shukur (2005), Alkhamisi et al. (2006), Alkhamisi and Shukur (2007), Muniz and Kibria (2009), Månsson et al. (2010), Muniz et al. (2012). In ridge regression, a small positive number k is added to the diagonal entries of the correlation matrix $X'X$ to decreases the variance so that one can obtain stable estimated coefficients. However, there is a disadvantage of the ridge estimator such that the estimated parameters are non-linear functions of the ridge parameter $k$ and that the small values of $k$ not be large enough to overcome multicollinearity.

Liu (1993) proposed the Liu estimator which has the advantage of being a linear function of the shrinkage parameter $d$. This estimator has the advantages of ridge estimator and Stein estimator Stein (1956).

Logistic regression, being a kind of generalized linear model, is one of the most widely used statistical analysis in many fields. Moreover, there is a growing interest in the problem of multicollinearity for the logistic regression model. We recommend the following studies: In Schaefer et al. (1984), a logistic ridge estimator was proposed for this problem. Månsson and Shukur (2011) and Kibria et al. (2012) proposed many ridge estimators to be used in logistic ridge regression. Månsson et al. (2012) proposed to generalize the Liu estimator to the binary logistic regression model. Huang (2012) also proposed a two-parameter estimator in logistic regression. Inan and Erdogan (2013) proposed the Liu-type logistic estimators by generalizing the Liu-type estimator (Liu, 2003) to the binary logistic regression model. Asar and Genç (2016) proposed some new shrinkage estimators for the Liu-type logistic estimator.

However, multicollinearity has not been studied in the multinomial logistic regression models so far. One exception is the paper of Månsson et al. (2018) in which the authors proposed to use ridge regression when there is a multicollinearity problem in the multinomial logistic regression. They conducted a Monte Carlo simulation study to evaluate the performance of the existing ridge estimators and the MLE using the mean squared error (MSE) criterion. Another exception is that Abonazel and Farghali (2019) proposed a two-parameter estimator including both ridge and Liu estimators in the multinomial logistic regression.

The purpose of this paper is to generalize the Liu estimator (Liu, 1993) to the multinomial logistic regression and propose some methods to estimate the shrinkage parameter $d$ used in multinomial logistic Liu estimator (MLLE) to combat multicollinearity in the multinomial logistic regression model. Liu estimator with the proposed shrinkage estimators is assumed to perform much better than the maximum likelihood estimation when the explanatory variables are correlated. We design a Monte Carlo simulation to evaluate the performance of the estimators. Mean squared error (MSE) is used to compare the estimators as a performance criterion.

This paper is organized as follows: In Section 2, we present the model, the methodology, and the proposed shrinkage estimators. The details of the Monte Carlo experiment are given in Section 3. We provide the results and discussions of the simulation in Section 4. An application of real data is demonstrated in Section 5. Finally, a summary and conclusion are presented.

## 2. Theory and Method

### 2.1. Maximum likelihood estimator

When the dependent variable consists of $m$ ($m > 2$) different categories, the multinomial logistic regression model developed by Luce (1959) is one of the most popular statistical methods. The multinomial logistic model can be specified as follows

$$\pi_{lj} = \frac{\exp(x_l \beta_j)}{\sum_{j=1}^{m} \exp(x_l \beta_j)} \tag{2.1}$$

where $x_l$ is the lth row of $X$ which is a $n \times (p+1)$ data matrix with $p$ explanatory variables and $\beta_j$ is a $(p+1) \times 1$ vector of coefficients. Most commonly, the method of maximum likelihood is used when estimating $\beta_j$ by maximizing the following log-likelihood function

$$L = \sum_{i=1}^{n} \sum_{j=1}^{m} y_{lj} \log(\pi_{lj}) \tag{2.2}$$

In order to maximize the Equation (2.2), we should obtain the derivative with respect to $\beta$ and equate the resulting function zero and then solve for $\beta$. We obtain the following subsequent equation:

$$\frac{\partial L}{\partial \beta_j} = \sum_{j=1}^{m} (y_{lj} - \pi_{lj}) x_i = 0. \tag{2.3}$$

Since these equations are nonlinear in $\beta$, the iteratively re-weighted least squares (IRLS) method is generally used as follows (Saleh and Kibria, 2013):

$$\hat{\beta}_j^{t+1} = \hat{\beta}_j^t + (X'\hat{W}_j^t X)^{-1} X'\hat{W}_j^t (y - \hat{\pi}_j^t) \tag{2.4}$$

where $\hat{W}_j^t = \text{diag}(\hat{\pi}_{lj}^t (1 - \hat{\pi}_{lj}^t))$ and where $\hat{\pi}_j^t$ is the estimated values of $\pi_j^t$ using $\hat{\beta}_j^t$ and $\hat{\pi}_{lj}^t$ is the lth element of $\hat{\pi}_j^t$. After some algebra, the Equation (2.4) becomes

$$\hat{\beta}_j^{MLE} = \left(X'\hat{W}_j^t X\right)^{-1} X'\hat{W}_j^t z_j \tag{2.5}$$

where $z_j' = (z_{1j} \cdots z_{nj})$ with $\eta_l = x_l' \beta_j$ and $z_{lj} = \eta_{lj} + (y_{lj} - \pi_{lj})(\partial \eta_{lj} / \partial \pi_{lj})$.

Månsson et al. (2018) obtained the asymptotic covariance and scalar mean squared error (MSE) of MLE respectively as follows:

$$\text{Cov}\left(\hat{\beta}_j^{MLE}\right) = \left(X'\hat{W}_j^t X\right)^{-1} \tag{2.6}$$

and

$$\text{MSE}\left(\hat{\beta}_j^{MLE}\right) = \sum_{i=1}^{p} \frac{1}{\lambda_{ji}} \tag{2.7}$$

where $\lambda_{ji}$ is the ith eigenvalue of $X'\hat{W}_j^t X$.

However, some of the $\lambda_{ji}$'s becomes close to zero such that the variance of $\hat{\beta}_j^{MLE}$ becomes inflated when there is multicollinearity between the explanatory variables. Due to this problem, $\hat{\beta}_j^{MLE}$ becomes unstable and has large standard errors.

### 2.2. Multinomial logistic Liu estimator

Since the variance of MLE is inflated, we propose to use the logistic Liu estimator (Månsson et al., 2012) to decrease the variance in the multinomial logistic model. Although Abonazel and Farghali (2019) mentioned slightly about the multinomial version of the Liu estimator, following Månsson et al. (2012), we define the multinomial logistic Liu (MLL) estimator as follows:

$$\hat{\beta}_j^{MLL} = \left(X'\hat{W}_j^t X + I_p\right)^{-1} \left(X'\hat{W}_j^t X + dI_p\right) \hat{\beta}_j^{MLE} \tag{2.8}$$

where $0 < d < 1$. It is easy to obtain the asymptotic covariance and bias of MLL as follows:

$$\text{Cov}\left(\hat{\beta}_j^{MLL}\right) = (C+I)^{-1}(C+dI)C^{-1}(C+dI)(C+I)^{-1} \tag{2.9}$$

and

$$\text{bias}\left(\hat{\beta}_j^{MLL}\right) = -(1-d)(C+I)^{-1}\beta_j \tag{2.10}$$

where $C_j = X'\hat{W}_j^t X$. Therefore, we can obtain the matrix mean squared error (MMSE) and scalar MSE of MLL respectively as follows:

$$\begin{aligned}\text{MMSE}\left(\hat{\beta}_j^{MLL}\right) &= E\left[\left(\hat{\beta}_j^{MLL} - \beta_j\right)\left(\hat{\beta}_j^{MLL} - \beta_j\right)'\right] \\ &= \text{Cov}\left(\hat{\beta}_j^{MLL}\right) + \text{bias}\left(\hat{\beta}_j^{MLL}\right)\text{bias}\left(\hat{\beta}_j^{MLL}\right)' \\ &= (C+I)^{-1}(C+dI)C^{-1}(C+dI)(C+I)^{-1} \\ &\quad + (1-d)^2 (C+I)^{-1} \beta_j \beta_j' (C+I)^{-1}\end{aligned} \quad (2.11)$$

and

$$\begin{aligned}\text{MSE}\left(\hat{\beta}_j^{MLL}\right) &= E\left[\left(\hat{\beta}_j^{MLL} - \beta_j\right)'\left(\hat{\beta}_j^{MLL} - \beta_j\right)\right] \\ &= \sum_{i=1}^{p} \frac{(\lambda_{ji}+d)^2}{\lambda_{ji}(\lambda_{ji}+1)^2} + (1-d)^2 \frac{\alpha_{ji}^2}{(\lambda_{ji}+1)^2}\end{aligned} \quad (2.12)$$

where $\alpha_{ji}^2$ is the square of the ith component of $\alpha_j = T_j' \beta_j$ such that $T_j$ is the matrix whose columns are eigenvectors of the matrix $X'\hat{W}_j^t X = T_j' \Lambda_j T_j$ and $\Lambda_j = \text{diag}(\lambda_{ji})$ such that $\lambda_{ji}$'s are the eigenvalues of $X'\hat{W}_j^t X$.

Månsson et al. (2012) showed that there is always a value of $d$ between zero and one so that $\text{MSE}\left(\hat{\beta}_j^{MLL}\right) < \text{MSE}\left(\hat{\beta}_j^{MLE}\right)$.

### 2.3. Estimators of the biasing parameter d

To estimate the parameter $d$, we differentiate the MSE function (2.12) with respect to $d$ and obtain the following function:

$$g'(d) = \sum_{i=1}^{p} \frac{2(\lambda_{ji}+d)}{\lambda_{ji}(\lambda_{ji}+1)^2} - 2(1-d)\frac{\alpha_{ji}^2}{(\lambda_{ji}+1)^2} . \quad (2.13)$$

where $g(d) = \text{MSE}\left(\hat{\beta}_j^{MLL}\right)$.

If we solve the Equation (2.13) for each parameter $d$, we obtain the following individual parameters:

$$d_{ji} = \frac{\lambda_{ji}\left(\alpha_{ji}^2 - 1\right)}{1 + \lambda_{ji}\alpha_{ji}^2}$$

Following Månsson et al. (2012), we propose to choose the biasing parameter $d_j$ as follows:

$$d_{j1} = \max\left(0, \frac{1}{p}\sum_{i=}^{p}\left(\frac{\lambda_{ji}\left(\alpha_{ji}^2 - 1\right)}{1 + \lambda_{ji}\alpha_{ji}^2}\right)\right),$$

$$d_{j2} = \max\left(0, \text{median}\left(d_{ji}\right)\right),$$

$$d_{j3} = \max\left(0, \min\left(d_{ji}\right)\right).$$

Since $\alpha_j^2$ is not known in reality, we estimate it by $\hat{\alpha}_j^2 = T'\hat{\beta}_j^{MLE}$.

## 3. Monte Carlo Simulation Experiment

### 3.1. Design of the simulation

In this section, we present the details and the results of the Monte Carlo simulation which is conducted to evaluate the performances of the estimators MLL and MLE. We refer to the following papers to design the simulation Lee and Silvapulle (1988), Asar and Genç (2016), and Månsson et al. (2012) generating explanatory variables as follows:

$$x_{li} = \left(1 - \rho^2\right)^{1/2} z_{li} + \rho z_{lp} \tag{3.1}$$

where $\rho^2$ is the correlation between the explanatory variables, $l = 1, 2, ..., n$, $i = 1, 2, ..., p$ and $z_{li}$ are random numbers generated from the standard normal distribution. Effective factors in designing the simulation are the number of explanatory variables $p$, the degree of the correlation among the independent variables $\rho^2$, and the sample size $n$.

Three different values of the correlation $\rho$ corresponding to 0.9, 0.99, and 0.999 are considered. Moreover, four different values of the number of explanatory variables consisting of 4, 8, 12, and 20 are considered in the design of the experiment. The sample size varies as 100, 200, 500, and 1000.

The coefficient vector $\beta_j$ is chosen due to Newhouse and Oman (1971) such that $\beta_j' \beta_j = 1$ which is a commonly used restriction, for example, see Kibria (2003). We generate the $n$ observations of the dependent variable using

$$\pi_{lj} = \frac{\exp(x_l \beta_j)}{\sum_{j=1}^{m} \exp(x_l \beta_j)} \quad (3.2)$$

where $l = 1, 2, ..., n$ and $j = 1, 2, ..., m$ $x_l$ is the lth row of the data matrix $X$.

The simulation is repeated 2000 times. To compute the simulated MSEs of the estimators, the following equation is used respectively:

$$\mathrm{MSE}(\tilde{\beta}) = \frac{\sum_{i=1}^{R} \sum_{j=1}^{m} (\hat{\beta}_j - \beta_j)'(\hat{\beta}_j - \beta_j)}{R} \quad (3.3)$$

where $\tilde{\beta}$ is MLE or MLL in the simulation and R is the number of replications. The convergence tolerance is taken to be $10^{-6}$.

### 3.2. Results of the simulation

The simulated MSE values of the estimators are reported in Tables 1-4. According to the tables, we can make the following conclusions:

- It is observed that increasing the degree of correlation affects the estimators negatively, namely, their MSE values increases. Especially, the MSE of MLE is inflated when the degree of correlation is increased. The least affected estimator is MLL with $d_3$ from this situation.

- Moreover, we observe that increasing the sample size makes a positive effect on the estimators. MSE of MLE decreases rapidly when the sample size increases.
- It can also be observed from tables that if we increase the number of explanatory variables, the estimators are affected negatively, especially, MSE of MLE is inflated.
- If we compare the estimators of the biasing parameter $d$, we see that $d_3$ has the best performance. Actually, MLL with $d_3$ has the best performance in all situations.
- Furthermore, we also observe that when the degree of correlation is 0.9 and the sample size is 500 or 1000, the performance of the estimators becomes close to each other. When p=4, MLL with $d_1$ is always better than MLL with $d_2$.
- However, if we increase $p$, we observe that increasing the degree of correlation or the sample size makes MLL with $d_2$ better than MLL with $d_1$. For example, when $p = 20$, although the performance of MLL with $d_1$ is better than $d_2$ for small sample sizes and $\rho = 0.9$, the performance of MLL $d_2$ becomes better than MLL with $d_1$ for large sample size (n $= 1000$) and $\rho = 0.999$.
- Overall, MLL with $d_3$ is the best estimator among the others. Therefore, we recommend using MLL in multicollinear situations in the multinomial logistic regression models.

**Table 1.** MSE values of the estimators when $p = 4$

| $\rho$ | n | 100 | 200 | 500 | 1000 |
|---|---|---|---|---|---|
| 0.9 | MLE | 7.6023 | 2.7658 | 1.5937 | 1.1544 |
| | $d_1$ | 2.4293 | 1.6282 | 1.3164 | 1.0756 |
| | $d_2$ | 2.8835 | 1.6895 | 1.3190 | 1.0756 |
| | $d_3$ | 2.1209 | 1.6100 | 1.3161 | 1.0756 |
| 0.99 | MLE | 65.2970 | 20.7964 | 8.8854 | 4.9233 |
| | $d_1$ | 12.6354 | 5.2262 | 3.1604 | 2.4067 |
| | $d_2$ | 15.7321 | 6.0795 | 3.6100 | 2.5744 |
| | $d_3$ | 3.6473 | 2.6777 | 2.4777 | 2.3293 |
| 0.999 | MLE | 621.8735 | 218.5666 | 85.5395 | 44.3507 |
| | $d_1$ | 110.1127 | 42.1070 | 18.4062 | 9.9579 |
| | $d_2$ | 142.3089 | 53.0292 | 22.8936 | 12.3244 |
| | $d_3$ | 22.4532 | 10.9331 | 5.9181 | 4.1250 |

**Table 2.** MSE values of the estimators when $p = 8$

| $\rho$ | n | 100 | 200 | 500 | 1000 |
|---|---|---|---|---|---|
| 0.9 | MLE | 19.9036 | 9.4234 | 3.2992 | 1.8575 |
| | $d_1$ | 2.4308 | 2.8674 | 2.2592 | 1.5862 |
| | $d_2$ | 2.6021 | 2.9838 | 2.2592 | 1.5862 |
| | $d_3$ | 2.3970 | 2.8646 | 2.2592 | 1.5862 |
| 0.99 | MLE | 260.6850 | 105.2291 | 27.6685 | 12.7491 |
| | $d_1$ | 21.3082 | 12.3643 | 5.7438 | 4.0598 |
| | $d_2$ | 20.3407 | 13.3226 | 6.4772 | 4.2265 |
| | $d_3$ | 2.1521 | 3.7502 | 4.9143 | 4.0367 |
| 0.999 | MLE | 2970.4591 | 1070.0242 | 292.5144 | 129.5360 |
| | $d_1$ | 279.4341 | 129.9201 | 38.4700 | 19.1291 |
| | $d_2$ | 267.2931 | 148.6973 | 51.4409 | 23.8003 |
| | $d_3$ | 6.9773 | 5.1838 | 4.0413 | 4.2304 |

**Table 3.** MSE values of the estimators when $p = 12$

| $\rho$ | n | 100 | 200 | 500 | 1000 |
|---|---|---|---|---|---|
| 0.9 | MLE | 102.1019 | 18.8089 | 5.4591 | 2.7511 |
| | $d_1$ | 3.4124 | 3.5680 | 3.1023 | 2.0904 |
| | $d_2$ | 3.8720 | 3.5857 | 3.1023 | 2.0904 |
| | $d_3$ | 3.0903 | 3.5680 | 3.1023 | 2.0904 |
| 0.99 | MLE | 1221.8327 | 205.7229 | 55.6312 | 23.9824 |
| | $d_1$ | 76.3228 | 15.8034 | 7.4737 | 5.9099 |
| | $d_2$ | 56.8848 | 14.3924 | 7.9590 | 6.0956 |
| | $d_3$ | 2.6480 | 4.2341 | 6.5757 | 5.8778 |
| 0.999 | MLE | 13000.5328 | 2230.8238 | 585.1818 | 254.3137 |
| | $d_1$ | 1026.3804 | 204.1690 | 61.3551 | 31.2063 |
| | $d_2$ | 776.1340 | 188.8036 | 64.5309 | 35.1376 |
| | $d_3$ | 5.4482 | 2.9191 | 3.8887 | 5.0751 |

**Table 4.** MSE values of the estimators when $p = 20$

| $\rho$ | n | 100 | 200 | 500 | 1000 |
|---|---|---|---|---|---|
| 0.9 | MLE | 259.1340 | 96.8769 | 10.6783 | 5.0480 |
| | $d_1$ | 6.1392 | 4.4151 | 4.3791 | 3.0956 |
| | $d_2$ | 7.5782 | 4.4710 | 4.3791 | 3.0956 |
| | $d_3$ | 5.0774 | 4.4120 | 4.3791 | 3.0956 |
| 0.99 | MLE | 2611.5978 | 899.6989 | 155.2784 | 57.5599 |
| | $d_1$ | 152.3736 | 54.9982 | 10.7373 | 7.8509 |
| | $d_2$ | 110.4739 | 43.2410 | 10.3712 | 7.9807 |
| | $d_3$ | 3.3608 | 5.3868 | 8.1111 | 7.8414 |
| 0.999 | MLE | 26539.3798 | 9813.9409 | 1611.5866 | 618.3079 |
| | $d_1$ | 2013.8061 | 858.4191 | 131.1522 | 54.7098 |
| | $d_2$ | 1524.1718 | 650.0063 | 111.5414 | 51.4790 |
| | $d_3$ | 2.7512 | 2.4219 | 4.2117 | 6.1792 |

## 4. A Real Data Application

In this section of the study, the data compiled for three different plant species from the leaf data set created by Silva (2013) was analyzed. Each plant's scientific name, the number of leaf specimens by species, and the texture features of leaves analyze are given in Table 5. The correlation matrix of the texture features of leaves is given in Table 6.

When the correlation matrix is examined, it is seen that there are high correlations between texture features. The square root of the ratio between the maximum and min eigenvalue of the correlation matrix for explanatory variables is called condition number (CN).

$$CN = \sqrt{\frac{\lambda_{max}}{\lambda_{min}}}$$

Since there are two weight matrices in the multinomial model, we compute the condition numbers of each matrix as CN1=384.845 and CN2=598.147. The values of CN between 10 and 30 indicate the presence of multicollinearity and when a value is larger than 30, the multicollinearity is regarded as strong (Kim, 2019). Therefore, we conclude that there is a strong multicollinearity problem with leaf data.

Table 5. Description of the leaf data set

| Response Variable | Level (Scientific Name) | Number of leaf specimens |
|---|---|---|
| Plant Species | Quercus suber | 12 |
| | Ouercus robur | 12 |
| | Betula pubescens | 14 |
| **Explanatory Variables** Texture features | **Descriptive** | |
| $X_1$ | Average intensity | |
| $X_2$ | Average Contrast | |
| $X_3$ | Smoothness | |
| $X_4$ | Third moment | |
| $X_5$ | Uniformity | |
| $X_6$ | Entropy | |

Table 6: Correlation matrix for texture features in the leaf data set

|       | $X_1$  | $X_2$  | $X_3$  | $X_4$  | $X_5$  | $X_6$  |
|-------|--------|--------|--------|--------|--------|--------|
| $X_1$ | 1.0000 | 0.9647 | 0.9821 | 0.9106 | 0.8609 | 0.9780 |
| $X_2$ |        | 1.0000 | 0.9828 | 0.9602 | 0.7292 | 0.9516 |
| $X_3$ |        |        | 1.0000 | 0.9714 | 0.7690 | 0.9455 |
| $X_4$ |        |        |        | 1.0000 | 0.6052 | 0.8672 |
| $X_5$ |        |        |        |        | 1.0000 | 0.8217 |
| $X_6$ |        |        |        |        |        | 1.0000 |

**Table 7.** Estimates and standard errors of the coefficients according to MLE and proposed Liu-type estimators

| | **Estimates of coefficient (standard error)** | | | |
|---|---|---|---|---|
| **Level: Ouercus suber** | MLE | $d_1$ (0.427) | $d_2$ (0.503) | $d_3$ (0.019) |
| $X_1$ | 41.13 (734.45) | 17.48 (313.93) | 20.59 (369.29) | 0.59 (13.64) |
| $X_2$ | -133.26 (78.91) | -57.07 (33.74) | -67.10 (39.68) | -2.66 (1.48) |
| $X_3$ | 226.26 (827.02) | 96.62 (353.50) | 113.68 (451.83) | 4.04 (15.36) |
| $X_4$ | -106.11 (301.00) | -45.44 (128.66) | -53.43 (151.34) | -2.12 (5.59) |
| $X_5$ | -46.28 (73.04) | -19.84 (31.23) | -23.32 (36.73) | -0.95 (1.38) |
| $X_6$ | -30.80 (159.72) | -13.30 (68.27) | -15.60 (80.31) | -0.80 (2.98) |
| MSE | 1351056.21 | 274696.90 | 362577.06 | 82317.46 |
| **Level: Ouercus robur** | MLE | $d_1$ (0.489) | $d_2$ (0.523) | $d_3$ (0.100) |
| $X_1$ | 561.76 (559.85) | 274.74 (273.74) | 293.98 (292.91) | 56.44 (56.13) |
| $X_2$ | -33.51 (39.22) | -16.29 (19.19) | -17.45 (20.53) | -3.20 (3.96) |
| $X_3$ | -876.17 (713.28) | -428.33 (348.76) | -458.34 (373.18) | -87.71 (71.51) |
| $X_4$ | 394.83 (280.20) | 193.14 (137.00) | 206.65 (146.60) | 39.73 (28.09) |
| $X_5$ | -25.97 (36.07) | -12.71 (17.65) | -13.60 (18.89) | -2.62 (3.64) |
| $X_6$ | -1.45 (109.80) | -0.60 (53.69) | -0.66 (57.45) | 0.04 (11.02) |
| MSE | 915604.86 | 542994.26 | 532750.28 | 1013788.07 |
| **Total MSE** | 2266661.08 | 817691.16 | 895327.34 | 1096105.53 |

We fit a multinomial logistic regression model using the IRLS algorithm. We compute MLE and MLL estimators with the proposed estimates of the biasing parameter of $d$. Estimated coefficients, their standard errors and estimated MSE values of MLE and the proposed Liu estimators are given in Table 7. According to Table 7, the standard errors and the MSE values of the proposed Liu estimators are smaller than the MLE.

Moreover, we plot the estimated MSE values for each level according to different values of the biasing parameter $d$ in Figure 1. The bar plot of the estimated MSE values for each level according to MLE and proposed Liu- estimators are given in Figure 2. According to this figure, it is readily observed that the multinomial Liu estimator has a lower MSE value for each level.

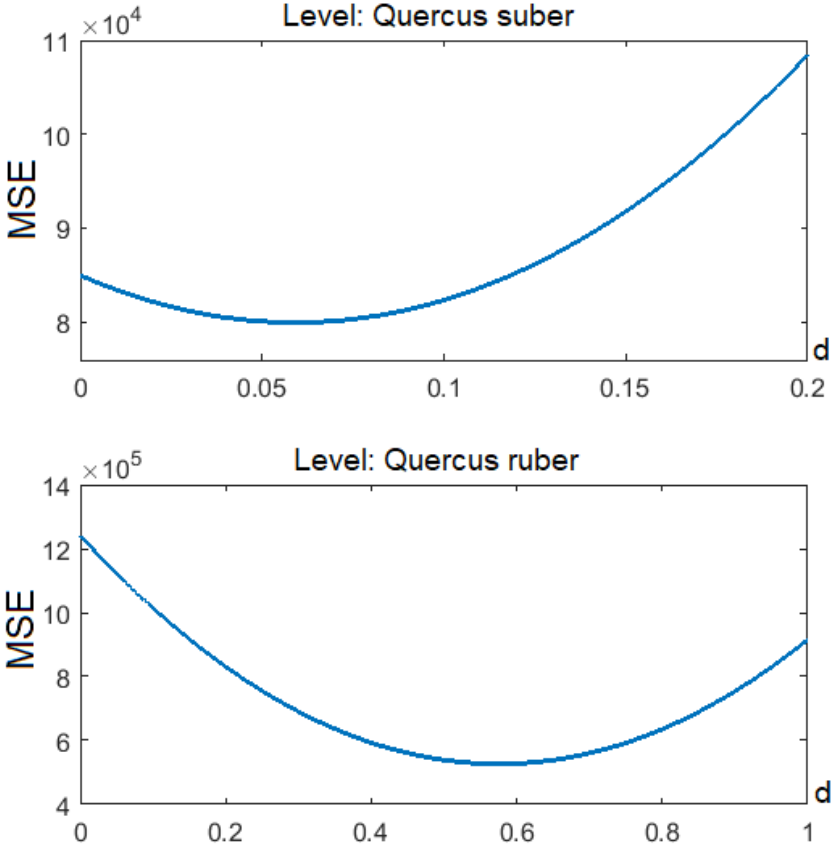

**Figure 1.** Estimated MSE values according to different values of the biasing parameter $d$

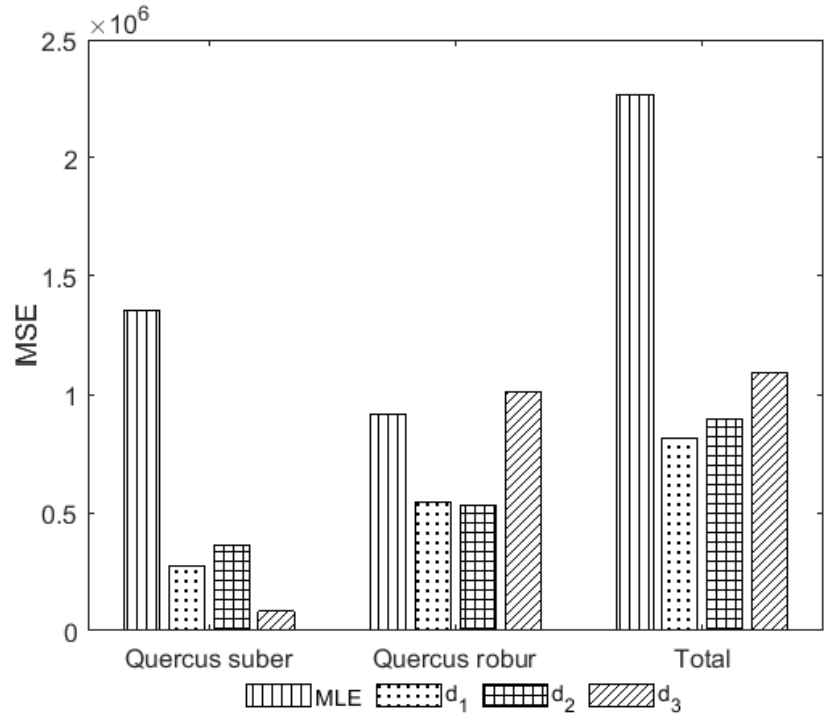

**Figure 2.** MSE estimations for MLE and proposed Liu-type estimators

## 5. Conclusion

This paper generalizes the well-known Liu estimator to the multinomial logistic regression model. We obtain the scalar mean squared error of the estimators MLE and MLL. Since there is biasing parameter $d$ in MLL, we propose some methods based on the study of Månsson et al. (2012) to estimate this parameter. A Monte Carlo simulation is designed to investigate the performance of MLL and MLE in the presence of different degrees of multicollinearity. Moreover, a real data example is presented to illustrate the findings of the paper. According to both the results of simulation and real application, we conclude that MLL has a better performance than MLE in multicollinear situations. Therefore, we recommend MLL to the researchers.

**Acknowledgments**: This paper was supported by Necmettin Erbakan University, Scientific Research Projects Unit, under Grant 161215004.